\date{July 1, 1999}
\documentstyle[preprint,eqsecnum,aps,epsfig]{revtex}
\def\lsim{\hbox{\lower .8ex\hbox{$\, \buildrel < \over \sim\,$}}}
\def\gsim{\hbox{\lower .8ex\hbox{$\, \buildrel > \over \sim\,$}}}
\tightenlines
\begin{document}
\draft
\title{Spectral spacing correlations for chaotic and disordered systems}
\author{ O. Bohigas$^{a}$, P. Leb{\oe}uf$^{a}$ and M. J. S\'anchez$^{b}$}
\maketitle
\noindent
\begin{center}
{$^{a}$\it Laboratoire de Physique Th\'eorique et Mod\`eles Statistiques
\footnote{Unit\'e de recherche de l'Universit\'e de Paris XI associ\'ee au
CNRS}, B\^at. 100, \\ Universit\'e de Paris-Sud, 91405 Orsay Cedex, France}
\center{$^{b}${\it Departamento de F\'{\i}sica J. J. Giambiagi, Facultad de
Ciencias Exactas y Naturales, \\ Universidad de Buenos Aires, Ciudad
Universitaria, 1428 Buenos Aires, Argentina.}}\\
\end{center}
\begin{abstract}
New aspects of spectral fluctuations of (quantum) chaotic and diffusive
systems are considered, namely autocorrelations of the spacing between
consecutive levels or spacing autocovariances. They can be viewed as a
discretized two point correlation function. Their behavior results from two
different contributions. One corresponds to (universal) random matrix
eigenvalue fluctuations, the other to diffusive or chaotic characteristics of
the corresponding classical motion. A closed formula expressing spacing
autocovariances in terms of classical dynamical zeta functions, including the
Perron-Frobenius operator, is derived. It leads to a simple interpretation in
terms of classical resonances. The theory is applied to zeros of the Riemann
zeta function. A striking correspondence between the associated classical
dynamical zeta functions and the Riemann zeta itself is found. This induces a
resurgence phenomenon where the lowest Riemann zeros appear replicated an
infinite number of times as resonances and sub-resonances in the spacing
autocovariances. The theoretical results are confirmed by existing ``data''.
The present work further extends the already well known semiclassical
interpretation of properties of Riemann zeros.
\end{abstract}

\vspace{4cm}

To appear in the Gutzwiller Festschrift, a special Issue of
Foundations of Physics.

\pagebreak


\section{Introduction}

Spectral fluctuations of classically chaotic or diffusive quantum systems have
been extensively studied in the past. One of the main outcomes of these
investigations has been to establish energy (or time) scales for which
universal properties hold and which are reproduced by random matrix theories
(RMT), to be distinguished from those which are system dependent and in many
cases can be treated by semiclassical theories \`a la
Gutzwiller$^{(\ref{gutz})}$. Apart from the one-point function or level
density $\rho$ which gives the global behavior and sets the main scale (its
average ${\bar \rho}$ defines the mean spacing ${\bar D} = {\bar \rho}^{-1}$,
or its conjugate variable defines the Heisenberg time $T_H = h {\bar \rho}$,
with $h$ the Planck constant), much emphasis has been put on the two-point
function $R_2 (\epsilon)$ or density-density correlation function or on
quantities therefrom derived ($R_2 (\epsilon) d\epsilon$ is proportional to
the probability of finding two levels separated by a distance in the interval
$[\epsilon,\epsilon+d\epsilon]$). One important exception to this is
constituted by the nearest neighbor spacing distribution $p(s)$ ($p(s)$ is not
a two-point function). It has been thoroughly investigated and its exact short
and long range behavior have simple analytical forms for the Gaussian
Ensembles. The short range behavior of $R_2 (\epsilon)$ at the scale of one
mean level spacing is the same as the one of $p(s)$, whereas the information
contained in the medium and long range part of $R_2 (\epsilon)$ (medium and
long range correlations) is not contained in $p(s)$. It is rather well
established that semiclassical theories are particularly adapted to describe
medium and long range correlations whereas their ability in dealing with short
range properties is dubious.

Studies of autocovariances $C(n)$ of the spacing between consecutive levels,
namely quantities related to the average value of $s_m s_{m+n}$, where $s_m =
t_{m} - t_{m-1}$, are scarce ($t_m$ denotes an eigenvalue, with $\ldots \leq
t_m \leq t_{m+1} \leq \ldots$). As will become clear in section II.A, spacing
autocovariances $C(n)$ can be viewed as a discrete version of the (continuous)
two-point correlation function (cf Eq.(\ref{ibn})). This is related to the
fact that the basic building block is a discrete object, a spacing between
consecutive levels. The discretization on a scale of the mean level spacing
induces a ``smoothing'' procedure, and the structures existing up to that
scale are strongly suppressed. As a consequence the spacing autocovariances
are particularly suited to be described by semiclassical approaches, which are
usually very difficult to control on the mean spacing scale. In terms of the
conjugate variable (time), the discretization amounts to introduce a cut-off
at the Heisenberg time $T_H$.

In contrast to $R_2 (\epsilon)$, exact analytical random matrix expressions
for the spacing autocovariances $C(n)$, whose study is one of the main
purposes of the present paper, are not known (this explains why they are
usually not extracted from data). In order to proceed use will be made of a
simple relation connecting spacing variances of distant eigenvalues to the
number variance (Eq.(\ref{onesix})). As discussed in Ref.${(\ref{bls1})}$,
this very general relation applies to eigenvalues of Gaussian Ensembles and
also to spectra of chaotic systems, even beyond the universal regime, as well
as to spectra of integrable systems (but does not hold for Poissonian
spectra). It is therefore associated to the concept of spectral rigidity,
characteristic of Gaussian Ensembles of random matrices and of the
non-universal regime of quantized Hamiltonian systems.

This paper, which may be considered as a continuation of Ref.${(\ref{bls1})}$,
is organized as follows. The next section starts with the general setting,
definitions and relations which will be subsequently used (section II.A). In
particular, we establish an approximate (but accurate) relation between
spacing autocovariances and the counting function variance (number variance).
When applied to eigenvalues of random matrices the resulting expressions are
consistent with an ansatz introduced in Ref.${(\ref{bls1})}$, though the
precise value of constants differ because of slight differences in both
treatments. Quantum systems governed by classically diffusive motion (section
II.B) and by chaotic dynamics (section II.C) are subsequently studied. In the
latter case the main tool used is a semiclassical approach. Because the
quantities computed are less sensitive to small energy scales, we find that
the simplest form of the theory, i.e. the ``diagonal approximation'', provides
a very accurate global description of $C(n)$. The main result of section II.C
is an expression of the autocovariances in terms of two different classical
dynamical zeta functions, one of them directly related to the spectrum of the
Perron-Frobenius operator. The structure of $C(n)$ can thus be interpreted as
a superposition of contributions of classical singular points (or resonances).
Section III has initially been conceived as an illustration of what precedes.
One studies properties related to the zeros of the Riemann $\zeta$-function by
interpreting them as eigenenergies of an hypothetical quantized classically
chaotic system, as suggested by Berry exploiting a beautiful analogy with the
Gutzwiller trace formula$^{(\ref{berry1})}$. Expressions for the spacing
autocovariances are worked out and compared to ``data'' provided by the
extensive computations of Odlyzko$^{(\ref{odlyzko})}$. Features encoded in
spacing autocorrelations are exhibited. Most of them are new and some
unsuspected. That is why we think that the content of the section has interest
in its own and goes beyond its initial illustrative purpose.

\label{sec:int}

\section{Spacing autocovariance}
\label{sec2}

\subsection{General Considerations}

Consider an ordered sequence of energy levels $t_i$, $i=1,2,\ldots$, having an
average density of states ${\bar \rho} (t)$. We call them ``energies'', but
the set can be an arbitrary sequence of points on the real line not
necessarily related to a quantum mechanical spectrum. From this sequence a new
stationary one, having mean level spacing equal to one, is constructed by the
unfolding or rectifying procedure$^{(\ref{bg}),(\ref{ls})}$ $x_i = \int^{t_i}
{\bar \rho} (t) {\rm d} t$. The sequence $x_i$ is located around energy $t$ in
a window of size $\Delta t$, with $\Delta t << t$.

Our purpose is to study properties related to distances between two
consecutive eigenvalues, denoted $s_m=x_m - x_{m-1}$. More precisely, we study
the autocovariances $C(n)$ of two spacings between consecutive eigenvalues
located $n$ levels apart
\begin{equation}\label{ij}
C(n) = \langle (s_{m} - \langle s_m  \rangle)(s_{m+n} -\langle s_{m+n}\rangle) = \langle s_{m}s_{m+n} \rangle - 1 = I(n) - 1  \ ,
\end{equation}
where $I(n)$ are the spacing autocorrelations. The brackets denote an average
over different locations $m$ of the reference spacing or, more generally, an
energy average over the spectrum. Eq.(\ref{ij}) is the simplest quantity
containing information on how the different spacings are correlated that can
be considered (correlations among spacings of non-consecutive levels can be
expressed in terms of $C(n)$ and do not contain new information). For
uncorrelated spacings $C(n) = \langle s_{m} \rangle \langle s_{m+n} \rangle -
1 = 0$.

Consider the length $S$ of an interval made of $n$ consecutive 
spacings, $S=\sum_{i=1}^{n} s_{i}\;$. The variance of $S$ can be written as
\begin{equation} \label{sigma2}
\sigma^2(n) = \; \langle S^{2} \rangle - {\langle  S \rangle}^2 =
\; n \left( \langle s^2 \rangle - n \right)  +
2\; \sum_{j=1}^{n-1} (n-j) \; I(j), \; \;  \; \; \;\; n \geq 2 \;,
\end{equation}
and $\sigma^2(1) = \langle s^{2} \rangle - 1$. Inversely, Eq.(\ref{sigma2})
allows to express the autocovariances in terms of the variance of $S$,
\begin{eqnarray} \label{isv}
C(n) & = & \frac{1}{2} \left[ \sigma^2 (n+1)  - 2\;  \sigma^2 (n) + \; 
\sigma^2 (n-1) \right] , \; \;  \; \; \;\; n \geq 2 \; , \\
\label{isvb}
C(1) & = & \frac{1}{2} \left[ \sigma^2 (2)  - 2\;  \sigma^2 (1) \right] 
 \; .
\end{eqnarray}
For eigenvalues of Gaussian Ensembles of random matrices neither $\sigma^2
(n)$ nor $C(n)$ are known in closed form. Our aim is to show that an 
approximate (but accurate) expression for the autocovariances of spacings
may be found exploiting some properties of $\sigma^2 (n)$ and Eq.(\ref{isv}).

For that purpose we consider the number variance $\Sigma^2 (L)$ closely
related to $\sigma^2 (n)$. It is defined as the variance of the random
variable which counts the number of levels contained in an interval of length
$L$ located randomly in the sequence. For eigenvalues of Gaussian Ensembles of
random matrices, $\Sigma^2 (L)$ evaluated at integer values of its argument is
related to the variance of the interval length through$^{(\ref{french})}$
\begin{equation} \label{onesix}
\Sigma^2 (L=n) - \sigma^2 (n) \approx 1/6 \ .
\end{equation}
This relation is valid, in principle, for large $n$. However, numerical
calculations as well as analytical estimates$^{(\ref{bls1})}$ indicate that
Eq.(\ref{onesix}) holds also for small values of $n$ with an error $\sim
0.01/n^2$. Eq.(\ref{onesix}) was used in Ref.${(\ref{bls1})}$ to determine the
coefficients of an ansatz for $C(n)$ in terms of a sum of inverse even powers
of $n$. In this section we follow an alternative path. We make no ansatz, but
use Eq.({\ref{onesix}) to derive similar random matrix theory (RMT)
expressions for the autocovariances $C(n)$. The next sections will be devoted
to the incorporation of the semiclassical and diffusive contributions to the
autocovariances.

Assuming that Eq.~(\ref{onesix}) holds, Eqs.~(\ref{isv})-(\ref{isvb}) can be
written 
\begin{eqnarray} \label{ibn}
C(n) & \approx & \frac{1}{2} \left[ \; \Sigma^2 (n+1)  - 2\;  \Sigma^2 (n)
 + \; \Sigma^2 (n-1) \; \right] \; \;\;\;\;\;\;\;\;\;\;\;\; n \geq 2 \ , \\
\label{ibnb}
C(1) & \approx & \frac{1}{2} \left[ \; \Sigma^2 (2)  - 2\;  \Sigma^2 (1)
 + \frac{1}{6} \; \right] \ . 
\end{eqnarray}
Due to the structure of Eq.(\ref{isv}) the exact value of the constant in the
r.h.s. of Eq.(\ref{onesix}) is immaterial for $n\geq 2$. Notice that
Eqs.(\ref{onesix}) and (\ref{ibn}) connect quantities related to the exact
location of energy levels, like $\sigma^2(n)$ or $C(n)$, to quantities which
are not, like $\Sigma^2$, and for which explicit results exist for chaotic and
diffusive systems$^{(\ref{berry2}),(\ref{mo})}$.

Using the approximation (\ref{ibn}) the autocovariances $C(n)$ are expressed
in terms of the (discrete) curvature of $\Sigma^2$ evaluated at integers $n$.
An analogous expression (up to a sign) relates for $L > 0$ the two-level
cluster function$^{(\ref{mehta1})}$ $Y_2 (L)$ to the (continuous) curvature of
$\Sigma^2 (L)$,
\begin{equation}\label{ys}
Y_2 (L) = - \frac{1}{2}\frac{\partial^2 \Sigma^2}{\partial L^2} \ .
\end{equation}
One therefore expects the properties of $C(n)$ to be closely related to the
ones of $Y_2$. One aim of the present paper is to identify some of their
differences.

Let us recall that the Fourier transform of $Y_2$ defines the form factor
$K(\tau)$, 
\begin{equation}\label{y2}
Y_2(L)= 2\; \int_0^\infty {\rm d} \tau (1 - K(\tau)) \; \cos(2 \pi L \tau)\ , 
\end{equation}
in terms of which the number variance reads
\begin{equation}\label{nv}
\Sigma^2 (L) = \frac{2}{\pi^2} \int_0^\infty {\rm d} \tau \ K(\tau) \
\frac{\sin^2 (\pi L \tau)}{\tau^2} \ .
\end{equation}
$\tau$ is a rescaled time in units of the Heisenberg time, $\tau = T/h {\bar
\rho}$ and, correspondingly, energies are being measured in units of mean
spacing ${\bar \rho}^{-1}$.

As is well known, the form factor $K(\tau)$ plays an important role in the
theory of quantum dynamical systems. For disordered and chaotic systems, it
contains basic information related to short time specificities. In the opposite
limit, namely the ergodic (long time) regime, its behavior is universal and
coincides with RMT. Accordingly, in order to exhibit the different relevant
time scales, it is convenient to write
\begin{equation}\label{s}
\Sigma^2  =  \Sigma^2_{\rm rm} \; + \Sigma^2_{\rm nu} \;.
\end{equation}
$\Sigma^2_{\rm rm}$ is the random matrix result for the number variance (cf
Eq.(\ref{srm2}) below). The non--universal contribution $\Sigma^2_{\rm nu}$
will be considered later on for disordered as well as for chaotic systems. From
Eq.(\ref{ibn}), and in analogy with Eq.({\ref{s}) we write
\begin{equation} \label{is}
C = C_{\rm rm}  + \; C_{\rm nu}.
\end{equation}

Using the exact form of $\Sigma^2_{\rm rm}$ it is possible to obtain, through
Eq.({\ref{ibn}), a general expression for $C_{\rm rm}$. However, we rather
prefer to use an asymptotic expansion of the number variance for large values
of $n$. In particular, this makes easier the comparison of the present
approach with the results of Ref.${(\ref{bls1})}$. For large values of $L$, the
leading order behavior of $\Sigma^2_{\rm rm}$ is$^{(\ref{mehta1})}$
\begin{equation}\label{s2a}
\Sigma^2 (L) = \frac{2}{\beta \pi^2} \log L + {\cal O}(1) \ ,
\end{equation}
where $\beta \;= 1, 2$ and $4$ denote the three types of Gaussian Ensembles,
orthogonal, unitary and symplectic, respectively. Eq.(\ref{s2a}), together
with Eq.(\ref{ibn}) leads to
\begin{equation}\label{cna}
C^0_{\rm rm} (n) = \frac{1}{\beta \pi^2} \log \left( 1 -
\frac{1}{n^2} \right)  \ ,
\end{equation}
valid for $n\geq 2$. For a more accurate evaluation of $C(n)$ we include more
terms in the asymptotic expansion of the number variance. For example, for
$\beta=2$ we use
\begin{equation}\label{srm2}
\Sigma^2 (L) = \frac{1}{\pi^2} \left[ \log(2\pi L) + \gamma + 1 -
\frac{\cos(2\pi L)}{(2\pi L)^2} - \frac{4\sin(2\pi L)}{(2\pi L)^3} + 
{\cal O}(1/L^4) \right] \ .
\end{equation}
An analogous expansion can be used for $\beta=1$. Substituting these
expansions in (\ref{ibn}) one gets, for $\beta =1$ and $2$
\begin{equation}\label{irmt}
C_{\rm rm} (n) = C^0_{\rm rm} (n) + 
\frac{1}{\beta \pi^2} \left( \frac{{\lambda}_{\beta}}{n^4} 
 \, + \frac{{\alpha}_{\beta}}{n^{6}} \right) + {\cal O}(1/n^8)  \; ,
\end{equation}
with ${\lambda}_{\beta}$ and ${\alpha}_{\beta}$ given by 
\begin{eqnarray} \label{i12}
{\lambda}_{1} & = & ~~ \frac{3}{2 \pi^2} \ , 
\;\;\;\;\; \alpha_{1}  =  ~~ \frac{15}{6 \pi^2} - \frac{105}{6\pi^4} \\ 
\nonumber \\
\label{i12b}
\lambda_{2} & = & - \frac{3}{2 \pi^2} \ , 
\;\;\;\;\; \alpha_{2}  = - \frac{15}{6 \pi^2} + \frac{135}{6 \pi^4} \ .
\end{eqnarray}
For $\beta=4$ a theorem relating the statistical properties of the Gaussian
Ensemble with $\beta = 1$ to those of $\beta= 4$ can be
exploited$^{(\ref{md})}$. It implies that the autocovariances are related as
follows (the notation is obvious),
\begin{equation}
\label{i4-1}
C_{\rm rm,4}(n)= \frac{1}{2} \;  C_{\rm rm,1}(2n) + \frac{1}{4} \; 
C_{\rm rm,1}(2n - 1) + \; \frac{1}{4} \; C_{\rm rm ,1}(2n +1) \; .
\end{equation}
Eq.(\ref{i4-1}) enables to determine the coefficients in (\ref{irmt}) for
$\beta=4$
$$
\lambda_{4} = \lambda_{1} /4 \ , \;\;\;\; 
{\alpha}_{4} = (\alpha_1 + 5 \; \lambda_{1})/16 \ .
$$

Aside from the logarithmic leading order term, Eq.(\ref{srm2}) contains
sub-dominant oscillatory corrections. The latter modify the simple estimate
Eq.(\ref{cna}) by adding smooth correction terms of order ${\cal O}(1/n^4)$
and higher according to Eq.(\ref{irmt}). Contrary to this, when computing the
cluster function from Eq.(\ref{ys}) these sub-dominant terms of $\Sigma^2 (L)$
give rise to an oscillatory behavior of $Y_2$ on the scale of the mean level
spacing. For example, for $\beta=2$ we have
\begin{equation}\label{y22}
Y_2(L)= \frac{1}{2 \pi^2 L^2} - \frac{\cos( 2 \pi L)}{2 \pi^2 L^2} \ .
\end{equation}
In contrast, there are no oscillations in $C(n)$ because $\Sigma^2$ is
evaluated at integer values of its argument, and the oscillatory functions in
Eq.(\ref{srm2}) are either zero or one. Similar results are also found for the
other symmetry classes. This makes an important difference between the RMT
behavior of $Y_2$ and that of $C(n)$, with a suppression of the structures of
$Y_2$ on the scale of the mean level spacing or below.

Expanding the logarithm in $C^0_{\rm rm} (n)$ one obtains from Eq.(\ref{irmt})
a series in inverse even powers of $n$, which is consistent with the ansatz of
Ref.${(\ref{bls1})}$. However the coefficients in this expansion slightly
differ from those obtained in Ref.${(\ref{bls1})}$. This difference is due to
the fact that terms of ${\cal O}(1/n^2)$ have been ignored in
Eq.(\ref{onesix}). Despite its asymptotic character, Eq.(\ref{irmt})
reproduces with good precision the autocovariances of spacings of random
matrices down to values $n\approx 2$.

Let us finally mention that the sum--rule
\begin{equation}\label{sumrule}
\sum_{j=1}^\infty C(j) + \frac{\sigma^2 (1)}{2} = 0
\end{equation}
valid for Gaussian Ensembles$^{(\ref{pandey})}$, is violated in the present
approach due to inaccuracies of the low ($j=1$ and $2$) terms ($C(1)$ has to
be computed from Eq.(\ref{ibnb})). This is in contrast with what is done in
Ref.${(\ref{bls1})}$ where, by construction, Eq.(\ref{sumrule}) is fulfilled.
Notice however that the procedure adopted here is much better adapted for a
semiclassical description and that this apparent deficiency is of no
consequence for what follows.

\subsection{Diffusive systems}
\label{dm}
We now consider the motion of a particle confined to a spatial region of
typical size $R$ whose classical dynamics is diffusive. The classical motion
is characterized by the diffusion constant $D$. The elastic mean free time
between elastic collisions is denoted $\tau_e = T_e /T_H$ (we again measure
the time in units of Heisenberg time $T_H = h {\bar \rho}$). Quantum
mechanically the relevant diffusive time scale is the Thouless time, i.e. the
typical time it takes to the system to cross the sample, $\tau_c = R^2/(D h
{\bar \rho})$ ($1/\tau_c = g$ is the dimensionless conductance). For a
diffusive system $\tau_e << \tau_c << 1$.

Ignoring the specificities of the dynamics for extremely short times $\tau
\lsim \tau_e$ $^{(\ref{ag})}$, for diffusive systems the form factor has, as
in the chaotic case, two distinct regimes. For times $\tau << \tau_c$ the
dynamics is diffusive (non-universal) and the form factor is given
by$^{(\ref{as})}$ 
\begin{equation}\label{kdif}
K_{\rm nu} (\tau) =
\frac{2}{\beta} \left( \frac{\tau_c}{4 \pi}\right)^{d/2} \tau^{1-d/2} \ ,
\end{equation}
with $d$ the space dimension. On the other hand, for $\tau >>\tau_c $ the
ergodic regime is reached and we recover a universal behavior for $K(\tau)$
described by RMT$^{(\ref{efetov})}$. We do not intend to describe in detail
the transition between these two regimes, and simply assume $K(\tau) = K_{\rm
rm}(\tau) + K_{\rm nu}(\tau)$ (consistently with Eq.(\ref{s})).

Inserting $K_{\rm rm} (\tau)$ in (\ref{nv}) we again get $\Sigma^2_{\rm rm}$
as in the previous subsection. On the other hand, from the diffusive
form factor (\ref{kdif}) we obtain $\Sigma^2_{\rm nu}$,
\begin{equation}\label{nvd}
\Sigma_{\rm nu}^2 (L) = \frac{4}{\beta \pi^2} \left(\frac{L \tau_c}{4\pi}
\right)^{d/2} \int_0^{\infty} {\rm d} \tau \ \frac{\sin^2 (\pi \tau)}
{\tau^{1+d/2}} \ .
\end{equation}
The integral gives a constant $a_d = \pi, \pi^2/2$ and $4\pi^2/3$ for $d=1, 2$
and $3$, respectively. Eq.(\ref{nvd}) provides the well known$^{(\ref{as})}$
asymptotic expression of the number variance for diffusive systems, $\Sigma^2
\propto (L\tau_c)^{d/2}$.

Computing the discrete curvature (\ref{ibn}) we have
\begin{equation} \label{isd}
C = C_{\rm rm}  + \; C_{\rm dif} \; .
\end{equation}
$C_{\rm rm}$ is the random matrix autocovariance given by (\ref{irmt}).
$C_{\rm dif}$ is computed from (\ref{nvd}),
\begin{equation} \label{id} 
C_{\rm dif}\ (n) = \frac{a_d \ d (d/2 - 1)}{\beta\pi^2} 
\left(\frac{\tau_c}{4\pi}\right)^{d/2}  
\frac{1}{n^{2-d/2}}  \ .
\end{equation}
Notice that for $d=2$ the contribution vanishes, and that it is positive for
$d=3$ and negative for $d=1$. Though this is a decreasing function of $n$, its
relative importance with respect to $C_{\rm rm}$ increases since for large
separations it vanishes as $n^{-(2-d/2)}$, which is slower than the $n^{-2}$
decay of $C_{\rm rm}$. It therefore dominates the tail of the autocovariance
function in any dimension (except $d=2$). It represents a finite conductance
correction to the RMT behavior, since as mentioned before $g = \tau_c^{-1}$.
For $d=3$, $C(n)$ is negative for small $n$ (spacings separated by small
distances are anti-correlated) but tends to zero from above for large
distances (positive correlations between large distant spacings (see Fig.~1)).
For $d=1$ $C(n)$ is negative and spacings are anti-correlated for any $n$.

\subsection{Chaotic systems}
\label{cm}
For ballistic fully chaotic systems $K(\tau)$ has two distinct
regimes$^{(\ref{berry2})}$. For times $\tau \approx \tau_{min}$, with
$\tau_{min} = T_{min}/h {\bar \rho}$ the rescaled period of the shortest
periodic orbit, the behavior of $K(\tau)$ is non-universal, with peaks located
at the periods of the shortest periodic orbits. For longer times the average
behavior of $K(\tau)$ is universal and consistent with RMT. The transition
between these two regimes is described semiclassically by the Hannay--Ozorio
de Almeida sum rule$^{(\ref{ho})}$, and occurs around a cut-off time $\tau_*$
satisfying $\tau_{min} << \tau_* << 1$. Correspondingly, for energies $n \lsim
1/\tau_*$, $\Sigma^2 (n)$ agrees with the random matrix prediction, while
around $n \approx 1/\tau_*$ there is a transition towards a non-universal
regime, where oscillations associated to short periodic orbits around a
saturation value occur. These are described$^{(\ref{berry2})}$ by the
non-universal term in (\ref{s})
$$
\Sigma^2_{\rm nu} = \Sigma^2_{\rm po} + \Sigma^2_{*} \ ,
$$
with
\begin{equation}\label{spo}
\Sigma^2_{\rm po} (n) = \frac{4}{\beta \pi^2} \; \sum_{r\tau_p<\tau_*}
 \frac{1}{r^2 |\det(M_{p}^r - 1)|}  \sin^2 (\pi n r\tau_p)
\end{equation}
and
\begin{equation} \label{ss}
\Sigma^2_{*}  =  \frac{2}{\beta \pi^2} \left[ {\rm Ci} (2 \pi n \tau_*) \; -
\log(2 \pi n\tau_*)  -\gamma \right] \; .
\end{equation}
In Eq.(\ref{spo}) the sum is over all the classical periodic orbits $p$ of the
system, with $r=1, 2, 3, \ldots$ the number of repetitions, $\tau_p = T_p
/T_H$ their normalized period and $M_p$ the mo\-nodromy (or stability) matrix.
The sum runs over orbits whose period (including repetitions) $r \tau_{p}$ is
smaller than $\tau_*$.

From (\ref{spo}) and (\ref{ss}) we get (cf (\ref{is}))
$$
C= C_{\rm rm} + C_{\rm nu}= C_{\rm rm} + C_{\rm po} + C_*
$$ 
with
\begin{equation} \label{ipo}
C_{\rm po} (n) = 
\frac{4}{\beta \pi^2} \sum_{r\tau_p<\tau_*} \frac{\sin^2 (\pi  r\tau_p)}
{r^2 |\det(M_{p}^r - 1)|}
\; \cos (2  \pi n r\tau_p) \; ,
\end{equation}
and
\begin{equation} \label{ist}
C_* (n) =  \frac{1}{\beta \pi^2} \left[{\rm Ci} (2 \pi (n+1) \tau_*)
-  2 {\rm Ci} (2 \pi n \tau_*) + {\rm Ci} (2 \pi (n-1) \tau_* ) 
- \log\left(1 - \frac{1}{n^2} \right) \right] \; . 
\end{equation}

The sum $C_{\rm po} + C_*$ accounts for the difference between the short time
dynamics of a real physical system and a pure RMT description. Indeed $C_*$
is, with opposite sign, the RMT contribution for times $0\leq\tau\leq\tau_*$
where a linear (diagonal) approximation of the form factor has been used. On
the other hand, due to the Hannay--Ozorio de Almeida sum rule, for times $\tau
\geq \tau_*$ the contribution of $C_{\rm po}$ coincides with $-C_*$. This
means that we may extend $\tau_*$ to infinity in $C_{\rm po}$ and $C_*$,
without affecting their sum. But from Eq.(\ref{ist}) we have $\lim_{\tau_*
\rightarrow \infty} C_* = -(\beta \pi^2)^{-1} \log (1 - 1/n^2)$. It follows
that
\begin{equation}\label{tsi}
C_{\rm po} + C_* = \frac{4}{\beta \pi^2} \sum_{p,r} 
\frac{\sin^2 (\pi  r\tau_p)}{r^2 |\det(M_{p}^r - 1)|}
\; \cos (2  \pi n r\tau_p) - \frac{1}{\beta \pi^2} \log \left( 1 -
\frac{1}{n^2} \right) \ .
\end{equation}
In contrast to Eq.(\ref{ipo}), there is no restriction imposed in the latter
equation and the sum extends over all periodic orbits $p$ and their
repetitions. Notice that the second term in the r.h.s. of Eq.(\ref{tsi})
cancels the term $C^0_{{\rm rm}}$ of the universal contribution $C_{\rm rm}$
in Eq.(\ref{irmt}). The autocovariance therefore takes the form (for $n\geq
2$)
\begin{equation}\label{cnf}
C (n) = \frac{4}{\beta \pi^2} \sum_{p,r} 
\frac{\sin^2 (\pi  r\tau_p)}{r^2 |\det(M_{p}^r - 1)|}
\; \cos (2  \pi n r\tau_p)  \, +
\frac{1}{\beta \pi^2} \left( \frac{{\lambda}_{\beta}}{n^4} 
\, + \frac{{\alpha}_{\beta}}{n^{6}} \right)   \, + {\cal O}(1/n^8)  \ .
\end{equation}
Eq.(\ref{cnf}) indicates that the essential features of the autocovariances
are described by the ``diagonal'' approximation given by the sum over the
periodic orbits, which includes the universal term $C_{\rm rm}^0$ given by
Eq.(\ref{cna}) as well as the non--universal fluctuations due to shorter
orbits\footnote{From a numerical point of view, instead of computing the sum
in Eq.(\ref{cnf}) over all the periodic orbits and its repetitions it is more
convenient to truncate the sum at some $r \tau_p < \tau_*$, where $\tau_*$ is
sufficiently large to ensure a RMT behavior, and add $-C_*$ (with $C_*$ given
by (\ref{ist})) to account for the remaining part of the sum.}. What is left
out of the sum are the smooth remaining RMT terms of order $1/n^4$ and higher
appearing in Eq.(\ref{cnf}), which are small compared to the leading order
behavior of $C_{\rm rm}^0 \approx -(\beta \pi^2 n^2)^{-1}$ because of the size
of the constants $\lambda_\beta$ and $\alpha_\beta$ (they are of order
$10^{-1}$), combined to the fact that $n\geq 2$ in Eq.(\ref{cnf}).

In summary, to a very good approximation the autocovariances $C(n)$ are given
by the first term in Eq.(\ref{cnf})
\begin{equation}\label{ipo3}
C (n) = \frac{4}{\beta \pi^2} \sum_{p,r} \frac{\sin^2 (\pi  r\tau_p)}
{r^2 |\det(M_{p}^r - 1)|} \; \cos (2  \pi n r\tau_p) \ .
\end{equation}
In this sum, the long, exponentially numerous periodic orbits produce a
coherent contribution given by Eq.(\ref{cna}), which is significant for low
values of $n$ and is independent of the particular position or height $t$ of
the window considered to compute $C(n)$ (i.e., is independent of $T_H$). On
the other hand, the amplitude of the contribution of the short orbits scales
like $\tau_p^2 \propto T_H^{-2}$. Therefore, in the extreme semiclassical
limit $T_H \rightarrow \infty$, only the universal term contributes. However,
at a finite $T_H$ the contribution of short trajectories must be taken into
account, and is particularly important for large values of $n$ where the
universal contribution is negligible. The distribution of the $\tau_p$, their
arithmetic and commensurability properties for low values of $p$, as
well as the stability of the orbits control the form of $C(n)$ for large
values of $n$. In the simplest (and probably general) case where only a few
non--correlated short orbits significantly contribute to $C(n)$, the sum
(\ref{ipo3}) leads to strong long range correlations between the spacings
$s_m$. The exact form depends on the interference pattern between the orbits,
with a rough estimate of the amplitude given by $4 \tau_{min}^2 /\beta
|\det(M_{min} - 1)|$. However, coherent interference contributions between
several orbits and its repetitions are of course possible, and larger
fluctuations cannot be excluded.

In order to clarify this point and to acquire further insight on the structure
of $C (n)$ we rewrite the sum (\ref{ipo3}) in a different manner. Our purpose
is to express the autocovariance in terms of classical zeta functions, a
procedure that leads to a more transparent analysis. For that end, we first
expand the sine function in a power series to get
$$
C (n) = \frac{1}{\beta} \sum_{k=1}^\infty (-1)^{k+1} \ \frac{\pi^{2(k-1)}2^{2k+1}}
{(2 k)!} \sum_{p,r} \frac{r^{2(k-1)} \tau_p^{2 k}}
{|\det(M_{p}^r - 1)|}  \; \cos (2  \pi n r\tau_p)  \; .
$$
It is convenient to split these sums in two parts $C^{(1)}$ and $C^{(2)}$
corresponding, respectively, to the two terms on the r.h.s. of the
decomposition of the $2 k$-th power of the period in the numerator
\begin{equation}\label{tp2}
\tau_p^{2 k} = r \ \tau_p^{2 k} - (r-1) \ \tau_p^{2 k} \ .
\end{equation}
The contribution of the first term to $C (n)$ may be written
$$
C^{(1)} (n) = - \frac{2}{\beta \pi^2} \sum_{k=1}^\infty \frac{1}{(2k)!}
{\rm Re} \frac{\partial^{2k}}
{\partial n^{2k}} \sum_{p,r} \frac{{\rm e}^{2 \pi i n r \tau_p}}
{r \ |\det(M_{p}^r - 1)|}
$$
or, alternatively
\begin{equation}\label{ipoz1}
C^{(1)} (n) = \frac{2}{\beta \pi^2} \sum_{k=1}^\infty \frac{1}{(2k)!} 
{\rm Re} \frac{\partial^{2k}}
{\partial n^{2k}} \log \prod_{p,r} \exp 
\left(- \frac{{\rm e}^{i r n T_p/\hbar
 {\bar \rho}}}{r \ |\det(M_{p}^r - 1)|} \right) \ .
\end{equation}
The argument of the $\log$ coincides with a well known classical dynamical
zeta function$^{(\ref{ce}),(\ref{rg})}$
\begin{equation}\label{zd}
Z (s) =  \prod_{p}\prod_{r=1} \exp \left(- \frac{{\rm e}^{s r T_p}}
{r \ |\det(M_{p}^r - 1)|} \right) \ ,
\end{equation}
and $C^{(1)}$ takes the form
\begin{equation}\label{ipoz}
C^{(1)} (n) = \frac{2}{\beta \pi^2} \sum_{k=1}^\infty \frac{1}{(2k)!} {\rm Re}
\frac{\partial^{2k}}{\partial n^{2k}} \log Z \left( i n/\hbar {\bar \rho}
\right) \ .
\end{equation}
This expression is similar, for $k=1$, to the relation found$^{(\ref{aaa})}$
between the diagonal part of the spectral two-point correlation function and
$Z(s)$ (see also the concluding remarks).

The contribution to $C(n)$ coming from the second term in the r.h.s. of
Eq.(\ref{tp2}), denoted $C^{(2)}$, can be obtained following similar
steps to those leading to Eq.(\ref{ipoz}). We obtain
\begin{equation}\label{ipoz2}
C^{(2)} (n) = - \frac{2}{\beta \pi^2} \sum_{k=1}^\infty \frac{1}{(2k)!} 
{\rm Re} \frac{\partial^{2k}}{\partial n^{2k}}
\log F \left( i n/\hbar {\bar \rho} \right) \ ,
\end{equation}
with $F$ another dynamical zeta function
\begin{equation}\label{fd}
F (s) =  \prod_{p}\prod_{r=2} \exp \left(- \frac{(r-1) {\rm e}^{s r T_p}}
{r^2 \ |\det(M_{p}^r - 1)|} \right) \ .
\end{equation}
By this procedure, the two terms $C^{(1)} (n)$ and $C^{(2)} (n)$ are expressed
without any approximation in terms of classical dynamical zeta functions.
Eq.(\ref{ipo3}) can thus be written in the compact form
\begin{equation}\label{cpoz}
C (n) = \frac{2}{\beta \pi^2} \sum_{k=1}^\infty \frac{1}{(2k)!} {\rm Re} 
\frac{\partial^{2k}}{\partial n^{2k}}
\log \left[ \frac{Z \left(i n/\hbar {\bar \rho}\right)}
     {F \left(i n/\hbar {\bar \rho}\right)} \right]  \ .
\end{equation}
By using the fact that for chaotic systems the periodic orbits are unstable
and that $r\geq 2$ in the expression of $F(s)$, it is reasonable to assume 
$|\det(M_{p}^r - 1)|\gg 1$. This leads to the approximate expression
\begin{equation}\label{fda}
F (s) \approx  \prod_{p}\prod_{r=2} \left(1 - \frac{{\rm e}^{s r T_p}}
{|\det(M_{p}^r - 1)|} \right)^{\frac{r-1}{r^2}} \ .
\end{equation}
Equation (\ref{cpoz}), together with Eqs.(\ref{zd}), (\ref{fd}) or
(\ref{fda}), is one of the main results of this paper. The net outcome has
been a transcription of the sum over periodic orbits by an expression in terms
of classical dynamical zeta functions. The first one, $Z(s)$, is well
known$^{(\ref{ce}),(\ref{rg})}$. Its complex zeros determine the spectrum
$\gamma_\mu$ of the Perron-Frobenius evolution operator of a classically
chaotic system. Physically, the characteristic times $({\rm Re}
\gamma_\mu)^{-1}$ determine the relaxation of the time evolution of classical
initial clouds. The second zeta function, $F(s)$, is formally very similar to
$Z(s)$ but has, to our knowledge, not been studied so far.

Assuming $Z(s)$ and $F(s)$ are entire functions that have no singularities
other than poles, the beauty of Eq.(\ref{cpoz}) is fully revealed by
considering the analytic structure of the classical zeta functions, i.e. their
singular points (zeros and poles) also called {\sl resonances}\footnote{By an
abuse of language, from now on we refer to poles {\sl and} zeros as
singular points. This terminology is justified because, up to a sign, from an
analytic function point of view they both play an equivalent role in
Eq.(\ref{cpoz}).}. To simplify the discussion, consider first the zeros
$\gamma_\mu$ of $Z(s)$ (other singular points are briefly discussed below).
Factorizing $Z(s)$ in terms of the $\gamma_\mu$, $Z(s) \propto \prod_\mu (s -
\gamma_\mu)$, and substituting this in Eq.(\ref{cpoz}), it follows that the
contribution to $C(n)$ of the zeros of $Z(s)$ is
\begin{equation}\label{ipog}
C (n) = \frac{1}{\beta \pi^2} \sum_\mu \log \left| 1 - \frac{1}
{(n + i \, n_\mu)^2} \right| \ ,
\end{equation}
where the $n_\mu$ are complex numbers determined by the (inverse) of the
classical times measured in units of Heisenberg time
$$
n_\mu = \hbar {\bar \rho} \ \gamma_\mu \ .
$$
The structure of $C (n)$ as a function of the real parameter $n$ may therefore
be viewed as resulting from the contributions of classical resonances located
at points $n_\mu$ in the complex plane. These are system dependent, except
$\gamma_0 = n_0 = 0$ which always exists for ergodic systems and brings in the
universal contribution Eq.(\ref{cna}). The remaining resonances $\gamma_\mu$,
$\mu \geq 1$ contribute to the non-universal oscillations of $C (n)$. Because
they are located at a finite position in the complex plane (different from the
origin), in the semiclassical limit $T_H \rightarrow \infty$ all the
non--universal resonances are pushed towards infinity ($|n_\mu| \rightarrow
\infty$ for $\mu \neq 0$), thus leaving the universal term $n_0 = 0$ as the
only remaining contribution. When $T_H$ is finite, each of the terms in
(\ref{ipog}) produces a peak in $C (n)$ whose shape is given by
\begin{equation}\label{fmu}
f_\mu = \frac{1}{2\beta \pi^2} \log \left[ 1 + \frac{2 n_{\mu R}^2 - 
2 (n - n_{\mu I})^2 + 1}{[n_{\mu R}^2 + (n - n_{\mu I})^2]^2} \right] \ ,
\end{equation}
where $n_{\mu R}$ and $n_{\mu I}$ are the real and imaginary part of $n_\mu$,
respectively. The peak is centered at $n \approx n_{\mu I}$, with a height $H
= (2\beta \pi^2)^{-1} \log [1 + 2(n_{\mu R}^2 + 1)/n_{\mu R}^4]$ and width $W
= 2 n_{\mu R}$. Away from the maximum, $f_\mu$ has a small negative tail that
tends to zero as $-[\beta\pi^2 (n-n_{\mu I})^2]^{-1}$. The position of the
peak is therefore controlled by the imaginary part $n_{\mu I}$, whereas the
real part $n_{\mu R}$ determines its height and width.

The other singular points of $Z(s)$ and $F(s)$ contribute with similar terms
as those of Eqs.(\ref{ipog}) and (\ref{fmu}). The only difference is a
possible overall sign with respect to the zeros of $Z(s)$. Due to the
structure of Eq.(\ref{cpoz}), the poles of $F(s)$ contribute with the same
sign as the zeros of $Z(s)$, whereas the zeros of $F(s)$ and the poles of
$Z(s)$ have the opposite sign. The final expression of $C(n)$ is therefore
given by a superposition of the contributions of all the singular points
$n_{\rm sp}$ of both functions, taking into account their appropriate sign,
\begin{equation}\label{ipof}
C (n) = \frac{1}{\beta \pi^2} \sum_{{\rm sp}} {\rm sign (sp)} \, \log \left| 1
- \frac{1} {(n + i \, n_{\rm sp})^2} \right| \ .
\end{equation}
The larger contributions will come from resonances with the smallest real
part.

\section{Application to the Riemann zeros}

We illustrate here the results of section II.C by considering the complex
zeros of the Riemann zeta function $\zeta (s) = \sum_n n^{-s}$. This function
is becoming an important model in the theory of quantized classically chaotic
Hamiltonian systems. The connexion with a dynamical system arises when the
complex zeros of $\zeta (s)$, located by the Riemann hypothesis on the line $s
= 1/2 + i \, t$, are interpreted as the eigenvalues of an hypothetical chaotic
system$^{(\ref{berry1}),(\ref{bk})}$.

At present there are many evidences supporting the idea that asymptotically
(i.e., for large $t$) the local statistical properties of the Riemann zeros
are described by the GUE distribution of random
matrices$^{(\ref{gm}),(\ref{odlyzko})}$, although more precise statements
concerning the appropriate matrix ensemble have recently been
made$^{(\ref{kzs})}$. In his extensive numerical studies of the statistical
properties of the zeros, Odlyzko$^{(\ref{odlyzko})}$ observed strong long
range correlations between the spacings $s_m$. He also found in the Fourier
transform of the autocovariances clear evidence of the contribution of small
prime numbers. Here we make a quantitative comparison between his computations
and the analytical results of the previous section.

For that purpose we adapt the equations of section II.C to the Riemann zeros
by exploiting the analogy between a classical dynamical system and the Riemann
zeta function. When comparing the Gutzwiller trace formula for the spectral
density of a chaotic system on the one hand and the density of zeros on the
critical line on the other (here interpreted as quantum eigenvalues), it
appears that the periodic orbits of the hypothetical ``Riemann dynamics'' are
labeled by the prime numbers $p = 2, 3, 5,\ldots$. Many properties of these
periodic orbits are known, and Table~I is a list of the more relevant ones.
In the same table we have also specified other quantities and correspondences,
like the (leading order) behavior of the density of zeros at a height $t$
along the critical axis.

With these definitions in hand, the autocovariance (\ref{ipo3}) reads
\begin{equation}\label{i3r}
C (n) =  \frac{2}{\pi^2} \sum_{p,r} \frac{\sin^2 (\pi r\tau_p)}
    {r^2 p^r} \; \cos(2 \pi n r \tau_p) \ .
\end{equation}
The sum now runs over the prime numbers $p$, and $\tau_p = \log p/\log
(t/2\pi)$ as indicated in Table~I. The parameter $t$ sets the height on the
critical line where the correlations are computed. In Figs.~2 and 3 the
theoretical prediction (\ref{i3r}) is compared with numerical values
(``data'') computed by Odlyzko. The agreement is good, although some small
deviations remain. Important differences with respect to the RMT result are
observed already at $n\approx 3$. For small values of $n$ (Fig.~2) the typical
amplitude of the non-universal fluctuations is of order $10^{-3}$. For larger
values of $n$ (Fig.~3), the amplitude of the oscillations is of order
$10^{-2}$, confirming the existence of strong long range correlations. Indeed,
assuming that the $s_m$ are uncorrelated variables, the statistical
fluctuations due to finite size effects are estimated$^{(\ref{odlyzko})}$ to
be of order $10^{-4}$. Besides their amplitude, Fig.~3 shows the existence of
sign correlations between several consecutive points.

It is instructive to look at the structure of $C(n)$ on larger scales
(Fig.~4). $C(n)$ is computed from Eq.(\ref{i3r}), which follows very closely
the numerical results of Odlyzko. For relatively small separations $50<n<500$,
$C(n)$ presents a remarkable behavior, with a dominant small positive
correlation ``punctuated'' by large peaks of anti-correlation (made of $3$ to
$4$ spacings). Qualitatively, the large peaks are produced by the constructive
interference of a relatively small number of short periodic orbits. Indeed, we
find that the qualitative features of the main resonances in Fig.~4(a) are
reproduced by summing in Eq.(\ref{i3r}) over approximately $12$ prime numbers
only (without repetitions). As $n$ increases the isolated peaks tend to
disappear, as well as the asymmetry of the plot with respect to the horizontal
axis (cf part (b) and (c) of Fig.~4). As we shall now see, this peculiar
structure finds a very simple explanation within the general framework based
on the classical zeta functions presented in section II.C.

The first step required to write $C(n)$ as a resonance formula corresponding
to Eqs.(\ref{cpoz}) and (\ref{ipof}) is to identify the classical dynamical
zeta functions associated to the ``Riemann dynamics''. This is achieved by
computing $Z(s)$ and $F(s)$ from Eqs.(\ref{zd}) and (\ref{fda}) using the
periodic orbit correspondences of Table~I (we use here for $F(s)$ the
approximate equation instead of the exact one). For the former one, this
procedure leads to
\begin{equation}\label{zr}
Z (s) = \prod_p \left( 1 - p^{s-1}\right) = \zeta^{-1} (1-s) \ ,
\end{equation}
where $\zeta (s) = \prod_p \left( 1 - p^{-s}\right)^{-1}$ is the Euler product
(over the prime numbers $p$) representation of $\zeta (s)$. Thus, when the
critical zeros of the Riemann zeta are interpreted as the eigenvalues
corresponding to the quantization of some classically chaotic motion, the
associated classical dynamical zeta function $Z(s)$ turns out to be the
inverse of the (translated) Riemann zeta itself. Concerning the other
dynamical function needed, $F(s)$, using again the periodic orbit
correspondences we have
$$
\prod_p \left( 1 - \frac{{\rm e}^{s r T_p}} 
{|\det(M_{p}^r - 1)|} \right) = \prod_p \left( 1 - p^{r s - r}
\right) = \zeta^{-1} \left( r - r s \right) \ ,
$$
and therefore, using the approximation (\ref{fda}) we obtain
\begin{equation}\label{fdr}
F (s) = \prod_{r=2} [\zeta (r - r s)]^{-(r-1)/r^2} \ .
\end{equation}
This shows the remarkable property that in the Riemann case the (approximate)
dynamical function $F(s)$ is also expressed in terms of $\zeta(s)$. The
dynamical zeta functions are basic tools to study the classical evolution. As
we have already emphasized, the Perron-Frobenius operator controls the time
evolution of classical statistical averages$^{(\ref{rg})}$. We have here
computed two of these classical functions for the ``Riemann dynamics'' which,
for completeness, have been added to Table~I.
 
The resulting expression for the autocovariances written in terms of dynamical
zeta functions follows from Eqs.(\ref{cpoz}), (\ref{zr}) and (\ref{fdr})
\begin{equation}\label{cpor}
C (n) = - \frac{1}{\pi^2} \sum_{k=1}^\infty \frac{1}{(2k)!} 
{\rm Re} \frac{\partial^{2k}}
{\partial n^{2k}} \log  \frac{\zeta ( 1 - i n/{\bar \rho} )}
{\displaystyle \prod_{r=2}^\infty [\zeta (r - i n r/{\bar \rho} )]^{(r-1)/r^2}}
 \ .
\end{equation}

To proceed further we shall exploit the analytic structure of $Z (s)$ and
$F(s)$. We simply need the following factorization formula$^{(\ref{tit})}$
\begin{equation}\label{fac}
\zeta (s) = \frac{{\rm e}^{(\log 2\pi - 1 - \gamma/2) s}}
{2 (s-1) \Gamma(1 + s/2)}
\prod_\kappa \left( 1 - \frac{s}{\kappa}\right) {\rm e}^{s/\kappa} \ ,
\end{equation}
which shows explicitly the presence of a pole at $s=1$, nontrivial
zeros $\kappa$ lying in the critical strip $0 \leq {\rm Re} \, \kappa \leq 1$,
as well as trivial zeros at the poles $s=-2m$, $m=1, 2, \ldots$ of $\Gamma(1 +
s/2)$. Assuming the Riemann hypothesis, namely $\kappa = 1/2 + i \, t_\mu$,
with $t_\mu$ real and coming in pairs symmetrical about the real axis, using
Euler's factorization formula for $\Gamma (1 + s/2)$, and deriving term by
term in the sums, Eq.(\ref{cpor}) leads, together with Eq.(\ref{fac}), to a
resonance-type formula
\begin{eqnarray}\label{ipogr}
C (n) = \frac{1}{2 \pi^2} \sum_{r=1} \left\{ \frac{(\delta_{r,1} + 1 - r)}{r^2}
\left[ \log \left|1 - \frac{1}{(n + i \, n_{0r})^2}\right| \right.\right. && -
\sum_\mu \log \left|1 - \frac{1}{(n + i \, n_{\mu r})^2}\right|     
\nonumber \\ 
&& \left.\left. - \sum_m \log \left|1 - \frac{1}{(n + i \, n_{mr})^2}\right|
\right] \right\} \ .
\end{eqnarray}
The outer sum in this equation is over the repetitions. Notice a global change
of sign (reflecting the structure of Eq.(\ref{cpor})) of the $r=1$ term with
respect to the $r\geq 2$ ones. The three terms inside the square brackets (of
which the last two are sums) originate from the pole, the critical zeros and
the trivial zeros of $\zeta(s)$, respectively. Due to the structure of
Eq.(\ref{cpor}) the singular points of $\zeta(s)$ appear in $C (n)$ as
resonances located at new, shifted positions in the complex plane given by
\begin{eqnarray}\label{res}
&& n_{0r} = {\bar \rho} \ (1 - 1/r) \nonumber \\
&& n_{\mu r} = {\bar \rho} \ [1 - (1/2 - i \, t_\mu)/r] \\
&& n_{m r} = {\bar \rho} \ [1 + 2 (m + 1)/r] \ . \nonumber 
\end{eqnarray}
Notice also, as mentioned in section II.C, that the contribution of the pole
has the opposite sign with respect to the terms associated to the zeros. 

Consider first the contributions of the singular points when $r=1$. Then the
resonances are located at $n_{01}=0$, $n_{\mu 1}={\bar \rho} \, (1/2 + i \,
t_\mu)$, and $n_{m1} = {\bar \rho} \, (2 m +3)$. The pole $n_{01}$ in
Eq.(\ref{ipogr}) gives the universal term Eq.(\ref{cna}). The nontrivial zeros
$n_{\mu 1}$ are located on a rescaled and shifted ``critical'' line, and have
all the same real part, ${\rm Re} \, n_{\mu 1} = {\bar \rho}/2$. This explains
the very interesting and peculiar structure of the autocovariances: for $r=1$
each nontrivial zero of $\zeta(s)$ contributes to $C (n)$ with a {\sl
negative} correlation peak whose shape is given by Eq.(\ref{fmu}), centered at
$n={\rm Im} \, n_{\mu 1} = {\bar \rho} \ t_\mu \, (= 55.1, 81.9, 97.4, \ldots$
for $\mu = 1, 2, 3, \ldots$, with ${\bar \rho} = 3.89533$ corresponding to the
window used by Odlyzko in his numerical computations). These negative peaks
are clearly visible in Figs.~\ref{fig4} and ~\ref{fig5}. The height $H$ and
width $W$ of the peaks is almost constant as observed in part $(a)$ of
Fig.~\ref{fig4} because the complex resonances are all at the same distance
${\bar \rho}/2$. From section II.C we have $H = (4 \pi^2)^{-1} \log [1 + 32
({\bar \rho}^2 /4 + 1)/ {\bar \rho}^4] \, (= 0.013)$ and $W=2 n_{\mu R} = {\bar
\rho}$, in good agreement with the heights and widths observed in
Fig.~\ref{fig4} and ~\ref{fig5}. The zeros $t_\mu$ are expected to produce
isolated peaks up to a value $n=n_c$ where their mean level spacing (measured
in units of ${\bar \rho} (t)$) becomes comparable to the width $W$ of the
peaks. This indicates the onset of a resonance overlap regime. This condition
gives $n_c \approx \log (t/2 \pi) \exp(2 \pi) \, (\approx 13100$ for the value
of $t$ used in the numerical calculations). On the other hand, the contribution
of the trivial zeros is small compared either to the complex zeros or to the
pole due to their greater distance with respect to the imaginary axis. The
curve with squares in Fig.~\ref{fig5} is a plot of the $r=1$ term of
Eq.(\ref{ipogr}) including all the singular points, compared to $C(n)$
computed from Eq.(\ref{i3r}). The presence of negative resonances at the
(rescaled) position of the lowest zeros was already noticed in the two-point
correlation function of the Riemann zeros in Ref.${(\ref{bk})}$ (see also the
concluding remarks).

Consider now the contributions of the singular points with $r\geq 2$, which
are at the origin of the smaller oscillations in Fig.~\ref{fig5}. Their
displacement in the complex plane as a function of $r$ is given by
Eq.(\ref{res}). In order to simplify the discussion, we consider explicitly
the critical zeros $n_{\mu r}$, but the analysis can be extended to the other
singular points. Two different effects on the location of these singular
points with respect to the $r=1$ distribution occur (see Eq.(\ref{res})). On
the one hand, as $r$ increases their real part increases, and ${\rm Re}\,
n_{\mu r} \rightarrow {\bar \rho}$ as $r\rightarrow\infty$. This implies that
the weight of their contribution to $C(n)$ decreases with increasing $r$. The
global factor $(1-r)/r^2$ in Eq.(\ref{ipogr}) goes in the same direction. On
the other hand, their location is compressed towards the real axis as $r$
increases, since ${\rm Im} \, n_{\mu r} = {\bar \rho} \, t_\mu /r$. Finally,
remember that the terms with $r\geq 2$ in (\ref{ipogr}) have the opposite sign
with respect to the $r=1$ terms.

We can therefore summarize the contribution of the critical zeros in
Eq.(\ref{ipogr}) as follows. Each critical zero $t_\mu$ produces a series of
peaks in the autocovariance. The main one has a {\sl negative} sign and is
located at $n = {\bar \rho} \, t_\mu$. Decorating the main resonance there is
a set of smaller {\sl positive} peaks located at $n={\bar \rho} \,
t_\mu/2,{\bar \rho} \, t_\mu/3,\ldots$ (with height diminishing as $H = (r-1)
(4 \pi^2 r^2)^{-1} \log [1 + 2 (({\rm Re}\, n_{\mu r})^2 + 1)/ ({\rm Re}\,
n_{\mu r})^4]$). Therefore, at a given value of $n$, the main contributions to
$C (n)$ come, with decreasing importance, from zeros located at $t_\mu =
n/{\bar \rho}, 2 n/{\bar \rho}, 3 n/{\bar \rho}, \ldots$. The smaller
oscillatory structure observed around a given $n$ is thus due to zeros located
$r=2, 3, \ldots$ times higher up the critical axis, thus producing an
interference pattern to which a large number of nontrivial zeros are
contributing with decreasing weight. The curve with diamonds in
Fig.~\ref{fig5} is a plot of Eq.(\ref{ipogr}) including up to the $r=3$ terms.
We observe that including terms higher than $r=1$ the agreement with the
result obtained from Eq.(\ref{i3r}) has greatly improved, and that the most
important features are reproduced. Even peaks up to the third generation are
easy to identify. For example, the small positive peak located at $n \approx
18.4$ corresponds to the third-order sub-resonance of the first Riemann zero
${\bar \rho} \, t_1/3$, with $t_1 \approx 14.13$. The resonance formula
Eq.(\ref{ipogr}) thus allows for a simple and transparent interpretation of
the structure of $C(n)$. In contrast to the periodic orbit sum Eq.(\ref{i3r}),
it allows to control the amount of complexity to be included 
{\sl hierarchically}.

Another interesting feature are the anomalously large peaks observed
occasionally in $C(n)$, as in part (b) of Fig.~\ref{fig4}. As mentioned
before, at the height $t$ fixed by the actual numerical computations each
complex zero of $\zeta(s)$ produces a peak in $C(n)$ of height $H\approx
0.013$ up to values $n \leq n_c$. However, this is true on average, and
statistical fluctuations of the spacings between consecutive zeros may produce
interference effects between neighboring peaks. In some extreme cases, we may
have two zeros which are very close to each other (on the scale of the width
of the peaks). The occurrence of zeros whose distance is much smaller than the
mean spacing is called Lehmer phenomenon, and is closely related to the
Riemann hypothesis$^{(\ref{odlyzko})}$. Whenever this unlikely phenomenon
between two zeros occurs, their contributions to $C(n)$ will add coherently
and produce a peak twice larger than the contribution associated to an
isolated zero. To illustrate this, we display in Fig.~\ref{fig6} the real
function
\begin{equation}\label{hardy}
{\cal Z} (t) = \exp({\rm i} \theta(t)) \ \zeta (1/2 + i \, t) \ ,
\end{equation}
with $\theta (t) = {\rm Im} \log \Gamma (1/4 + {\rm i} t/2) - (t \log \pi)/2$.
The presence of two almost degenerate zeros at $t = 7702/{\bar \rho} =
1977.24$ can be observed. This explains the large peak in $C(n)$ of size
$H\approx 0.026$ observed at $n=7702$ in part (b) of Fig.~\ref{fig4}.

The remarkable fact that, as for the optical image produced by two mirrors
face to face, the low-lying zeros of the Riemann zeta function emerge as a
decaying infinite sequence of replications when looking at the correlation of
zeros located at an arbitrary height $t$ on the critical axis appears more as
a conspiracy than as a generic feature of a dynamical system, although
this deserves further study. The origin of this singular property seems to be
associated to three (apparently unrelated) peculiarities of the Riemann zeros
which conspire to produce this effect: (i) the trace formula for their density
is exact, (ii) the period of the periodic orbits, $T_p = \log p$, does not
depend on energy (energy corresponds here to the location $t$ on the critical
axis), and (iii) the monodromy matrix has only one expanding eigenvalue.

\section{Summary and Conclusions}

In the present paper autocovariances $C(n)$ of two spacings between
consecutive levels located $n$ levels apart ($n=1,2,\ldots$) are studied.
Whereas the two-point correlation function $R_2 (\epsilon)$ can be viewed as
the curvature of the number variance $\Sigma^2$, $C(n)$ represents (to within
a sign) a discrete version of it. This fact implies that the effects present
on the form factor (Fourier transform of $R_2$) on the scale of the Heisenberg
time $T_H$ are absent on $C(n)$ (or on its Fourier transform). In particular,
we find that the non-analytic oscillatory structure of the RMT result of $R_2
(\epsilon)$, usually associated to off-diagonal contributions in semiclassical
theories, are absent in $C(n)$ which is, in the universal regime, a monotonic
function. The off-diagonal semiclassical contributions are here expected to
reproduce the smooth higher-order corrections written in Eq.(\ref{cnf}).

We first derive expressions for the autocovariance within the RMT of
Wigner-Dyson. They are approximate but accurate. In particular, it is shown
that the autocovariances in RMT are negative, monotonic functions of $n$,
and that for large $n$ they vanish like $- 1/\beta \pi^2 n^2$. The expressions
found can be compared to those obtained using a similar but alternative route
proposed in a previous paper. Both methods have their own advantages and
drawbacks. The method adopted here allows to express systematically selected
information on the form factor in terms of covariances. This is first
illustrated for the case of quantum systems which are classically diffusive.
An explicit expression of the autocovariances is given which consists of a
universal (random matrix) contribution plus a diffusive term which depends on
the dimensionality $d$ of the system. The latter, which appears to be
identically zero for two-dimensional systems, dominates the large $n$ behavior
of the autocovariances and has opposite sign for $d=1$ and $d=3$. Large
distant spacings are negatively correlated (anti-correlated) in one dimension
and positively correlated in three dimensions. Finally, for any dimension if
one neglects the finest time scale given by the elastic mean free time,
autocovariances tend to zero in the large $n$ limit.

Section II.C (study of quantum systems which are classically chaotic) and
section III (application to the zeros of the Riemann $\zeta$-function)
constitute the main body of the paper. By using periodic orbit theory we
arrive at an accurate formula for the autocovariances $C(n)$ in terms of a
diagonal approximation (Eq.(\ref{ipo3})). The important time scales are the
Heisenberg time (which is taken as unity) and the period of a short
periodic orbit $\tau_{min}$. Based on this result, we then find an expression
of the autocovariances in terms of a sum over derivatives of the usual
(classical) dynamical zeta function $Z(s)$ and another (classical) zeta
function $F(s)$ closely related to it (Eq.(\ref{cpoz})). The interpretation of
the singular points of $Z(s)$ is well known (relaxation of an initial
classical cloud, as given by the eigenvalues of the Perron-Frobenius
operator). In contrast, we are not aware of a similar discussion of $F(s)$.
Its study remains an open problem, in particular its analytic structure and
interpretation. Presumably that as far as autocovariances (and the two-point
correlation function) are concerned, $Z(s)$ gives the main behavior and the
role of $F(s)$ is limited to provide some additional finer structure. At any
rate, it is the location in the complex plane of the singular points (zeros
and poles) of $Z(s)$ and $F(s)$ that determine the structure of $C(n)$ through
a resonance-type formula (Eq.(\ref{ipof})). The location of the classical
singular points are rescaled by the Heisenberg time $T_H = h {\bar \rho}$. For
ergodic systems, in the semiclassical limit $T_H \rightarrow \infty$, only the
zero of $Z(s)$ at the origin gives a non-negligible contribution and $C(n)$
takes the universal form (\ref{cna}). At finite $T_H$ the non-universal
singular points contribute significantly.

The expression of $C(n)$ in terms of classical zeta functions is easy to
extend to the two point correlation function $R_2(\epsilon) = 1 - Y_2
(\epsilon)$. Using Eq.(\ref{y2}) as starting point, and by similar steps as
before we obtain, instead of Eq.(\ref{cnf}), the result
\begin{equation}\label{r2}
R_2 (\epsilon) - 1 = \frac{4}{\beta} \sum_{p,r} 
\frac{\tau_p^2}{|\det(M_{p}^r - 1)|}
\; \cos (2  \pi \epsilon r\tau_p)  \, + (R_2^{RMT} (\epsilon) - 1) + 
\frac{1}{\beta \pi^2 \epsilon^2}  \ ,
\end{equation}
where we are assuming an average over a small energy window that kills some
highly oscillatory frequencies in the sum over periodic orbits. $R_2^{RMT}$ is
the usual two-point function of RMT. As mentioned in Section II.A, the random
matrix correction to the diagonal sum involves a non-analytic oscillatory
behavior contained in $R_2^{RMT}(\epsilon)$ (cf Eq.(\ref{y22})), and which is
absent in $C(n)$. Due to this and contrary to the autocovariances of spacings,
the diagonal approximation is a poorer approximation for $R_2$. The term
$(\beta \pi^2 \epsilon^2)^{-1}$ plays here the role $-C_{\rm rm}^0 (n)$ is
playing for the autocovariances, and compensates the smooth large-$\epsilon$
decaying behavior of $R_2^{RMT}(\epsilon)$. The other important difference
between (\ref{r2}) and Eq.(\ref{cnf}) is the presence of the square of the
sine function in the latter, which is due to the discrete nature of the
curvature $C(n)$.

Exactly as for the autocovariances we may express the diagonal sum in
Eq.(\ref{r2}) in terms of classical dynamical zeta functions
\begin{equation}\label{r2zd}
R_2^{diag} (\epsilon) - 1 = \frac{1}{\beta \pi^2} {\rm Re}
\frac{\partial^{2}}{\partial \epsilon^{2}} \log \left[ \frac{Z \left(i
\epsilon/\hbar {\bar \rho}\right)} {F \left(i \epsilon/\hbar {\bar
\rho}\right)} \right] \ .
\end{equation}
The relation connecting the diagonal part of $R_2$ and $Z(s)$ was found in
Ref.${(\ref{aaa})}$ (where an extension to include the non-diagonal part was
also given, see also Ref.${(\ref{bok})}$). Here we have generalized this by
including a second dynamical zeta function, which we have explicitly computed
in the case of the Riemann $\zeta$-function and turns out to bring
non-negligible contributions. Its contribution was usually expressed as a sum
over periodic orbits, and not written in terms of the more transparent
approach used here. In contrast to Eq.(\ref{r2zd}), the existing relation
between $C(n)$ and the dynamical zeta functions Eq.(\ref{cpoz}) involves a sum
over derivatives, whose first term coincides with Eq.(\ref{r2zd}). The
summation leads to different shapes of the resonances of $C(n)$ compared to
those of Eq.(\ref{r2zd}).

These two quantities -- the spacings autocovariance $C(n)$ and the spectral
two-point correlation $R_2 (\epsilon)$ -- illustrate the manifestation of
classical resonances in quantum correlations. This connexion seems however to
be more general, as shown by a recent experiment$^{(\ref{pls})}$ on microwave
scattering in an open system.

We finally study, in the framework developed in the previous section on
chaotic systems, the autocovariances $C(n)$ for the case of zeros on the
critical line of the Riemann $\zeta$-function (section III). This is achieved
by using the well known ``dictionary'' which translates quantum (eigenvalues
of the Hamiltonian or evolution operator) and classical (periodic orbits,
actions,...) information into the corresponding $\zeta$-function quantities,
namely imaginary part of zeros (``quantum'') and prime numbers
(``classical'') (cf Table~I).

We first derive a periodic orbit sum (over primes and its repetitions)
expression of the autocovariances $C(n)$. After a short universal random
matrix regime, $C(n)$ is dominated with increasing $n$ by the non-universal
terms, encoded here in the prime numbers. We check the accuracy of the formula
obtained by comparing to partial existing ``data'' of Odlyzko. The agreement
is good. It is unclear whether the remaining differences are due to the
approximations made in our derivation or to numerical inaccuracies. Probably
to both. This deserves further study.

Better insight can be gained by writing the autocovariances in terms of the
(classical) dynamical zeta functions $Z(s)$ and $F(s)$. Extending the existing
``dictionary'' which connects dynamical properties and $\zeta$-function
properties, one finds that the classical dynamical zeta functions $Z(s)$ and
$F(s)$ are expressed in terms of the Riemann $\zeta$-function itself
(Eqs.(\ref{zr}) and (\ref{fdr}), and Table~I). This remarkable connection
between classical operators and the function that determines the ``quantum
spectrum'' leads to a resurgence phenomenon of the low lying Riemann zeros in
the autocovariances computed at a height $t>>1$ on the critical line. This
phenomenon is clearly displayed in a resonance formula obtained for $C(n)$
(Eq.(\ref{ipogr})) which is particularly appealing and which allows to
interpret most of the features observed from data as predicted by the periodic
orbit expression.

The salient qualitative and quantitative features of the autocovariances
$C(n)$, mentioned in what follows, are easy to read off from the resonance
formula. They have been checked by comparing to the results obtained from the
periodic orbit summation formula, which is accurate but comparatively opaque.
At a given height $t$ on the critical line, which sets the scale ${\bar
\rho}$, $C(n)$ starts as the random matrix prediction. Up to a critical value
$n_c$ ($\propto \log t$) of $n$, the low lying zeros of $\zeta$ produce
(negative) well isolated constant-amplitude peaks centered at $ {\bar \rho} \,
t_\mu$, where $t_\mu$ is the imaginary part of the $\mu$-th Riemann zero. One
therefore neatly sees the {\sl low-lying zeros} of $\zeta$ appearing in the
autocovariances computed at an arbitrary height on the critical line. The
effect of the other dynamical zeta function, $F(s)$, is to provide further
structure, namely smaller (positive) peaks at values $ {\bar \rho} \,
t_\mu/2$, $ {\bar \rho} \, t_\mu/3$, ..., with decreasing importance. One
therefore sees the low-lying zeros lurking as sub-resonant (positive) peaks.
Beyond $n_c$, the overlapping resonance regime sets in, and $C(n)$ oscillates
erratically. Occasionally, when two zeros of $\zeta$ are very close (Lehmer
phenomenon) at a height $t$, they give rise to a peak with double amplitude at
$n = {\bar \rho} \, t$.

Let us finally conclude by noticing that in several respects it is easier to
apply the formalism developed in Section II to the Riemann case, as we have
done here, than to a real dynamical system. It may be worth to study in some
detail some of the best known maps, like the cat or the baker map, from the
present viewpoint. The physical significance of the dynamical zeta function
$F(s)$ introduced here may then become clearer.

\vspace{0.8cm}

Acknowledgments: We are grateful to A. Odlyzko for discussions and
to the ECOS-Sud Franco-Argentinian scientific cooperation program (A98E03) for
partial financial support.

\pagebreak

\section*{REFERENCES}
\begin{enumerate}
\itemsep 0.01in
\item \label{gutz} M. C. Gutzwiller,  {\it Chaos in Classical and Quantum
    Mechanics} (Springer Verlag, New York, 1990).
\item \label{bls1} O. Bohigas, P. Leboeuf and M. J. S\'anchez. {\sl Physica D}
    {\bf 131}, 186 (1999).
\item \label{berry1} M. V. Berry, in {\sl Quantum Chaos and Statistical Nuclear
  Physics} edited by T. H. Seligman and H. Nishioka, {\it Lectures Notes in
  Physics} {\bf 263}, (Springer Verlag, Berlin, 1986) p.1. 
\item \label{odlyzko} A. M. Odlyzko, {\sl Math. Comp.} {\bf 48}, 273 (1987);
    {``The $10^{20}$-th zero of the Riemann zeta function and $70$ million of
    its neighbors''}, AT \& T Report, 1989 (unpublished); 
    {``The $10^{20}$-th zero of the Riemann zeta function and $175$ million of
    its neighbors''}, AT \& T Report, 1992 (unpublished).
\item \label{bg} O. Bohigas and M.-J. Giannoni, in {\sl Lecture Notes in
    Physics} {\bf 209}, (Springer, Berlin, 1984) p.1.
\item \label{ls} P. Leboeuf and M. Sieber, {\sl Phys. Rev. E} {\bf 60}, 3969
    (1999). 
\item \label{french} J. B. French, P. A. Mello and A. Pandey, {\sl Ann. Phys.}
     {\bf 113}, 277 (1978). T. A. Brody, J. Flores, J. B. French, P. A. Mello,
     A. Pandey and S. S. M. Wong, {\sl Rev. Mod. Phys.} {\bf53}, 385 (1981);
     J. B. French, V. K. B. Kota, A. Pandey and S. Tomsovic, {\sl Ann. Phys.}
     {\bf 181}, 198 (1988); A. Pandey, unpublished (1982).
\item \label{berry2} M. V. Berry, {\sl Nonlinearity} {\bf 1}, 399 (1988).
\item \label{mo} G. Montambaux in {\sl Quantum Fluctuations} edited by S.
    Reynaud, E. Giacobino and J. Zinn-Justin, {\it Les Houches Session LXIII},
    (North Holland, Amsterdam, 1997) p.387.
\item \label{mehta1} M. L. Mehta, {\it Random Matrices}
    (Academic Press, New York, 1991), 2nd ed. 
\item \label{md} M. L. Mehta and F. J. Dyson, {\sl J. Math. Phys.} {\bf 4},
    713 (1963).
\item \label{pandey} A. Pandey, in Ref.(\ref{berry1}), p.98.
\item \label{ag} A. Altland and Y. Gefen, {\sl Phys. Rev. Lett.} {\bf 71}, 3339
  (1993).
\item \label{as} B. L. Altshuler and B. I. Shklovskii, {\sl Zh. Eksp. Teor.
   Fiz.} {\bf 91}, 220 (1986) [{\sl Sov. Phys.} JETP {\bf 64}, 127 (1986)].
\item \label{efetov} K. Efetov, {\sl Adv. Phys.} {\bf 32}, 53 (1983).
\item \label{ho} J. Hannay and A. M. Ozorio de Almeida, {\sl J. Phys. A} {\bf
  17}, 3429 (1984).
\item \label{ce} P. Cvitanovi\'c and B. Eckhardt, {\sl J. Phys. A} {\bf 24},
  L237 (1991).
\item \label{rg} D. Ruelle, {\it Statistical Mechanics, Thermodynamic
  Formalism} (Addison-Wesley, Reading, MA, 1978); P. Gaspard, {\it Chaos,
  Scattering and Statistical Mechanics} (Cambridge University Press, Cambridge
  UK, 1998).
\item \label{aaa} O. Agam, B. L. Altshuler and A. Andreev, {\sl Phys. Rev.
  Lett.} {\bf 75}, 4389 (1995).
\item \label{bk} M. V. Berry and J. Keating, {\sl SIAM Review} {\bf 41}, 236
  (1999). 
\item \label{gm} H. L. Montgomery, {\sl Proc. Symp. Pure Math.} {\bf 24}, 181
  (1973); D. A. Goldston and H. L. Montgomery, {\sl Proc. Conf. at
  Oklahoma State Univ.}, edited by A. C. Adolphson {\sl et al}, 183 (1984).
\item \label{kzs} N. M. Katz and P. Sarnak, {\sl Bull. Amer. Math. Soc.} {\bf
  36}, 1 (1999).
\item \label{tit} E. C. Titchmarsh, {\it The Theory of the Riemann Zeta
  Function} (Clarendon Press, Oxford, 1986), $2$nd edition.
\item \label{bok} E. Bogomolny and J. Keating, {\sl Phys. Rev. Lett.}
  {\bf 77}, 1472 (1996).
\item \label{pls} K. Pance, W. Lu and S. Sridhar, {\sl Phys. Rev. Lett.}
  {\bf 85}, 2737 (2000).
\end{enumerate}

\begin{center}
\begin{table}[b]
\begin{minipage}{10cm}
\begin{tabular}{|l|l|}
            & Correspondences \\
\hline
~label of periodic orbits~ & prime numbers $p$ \\
\hline
~Planck constant & $\hbar \rightarrow 1$ \\
\hline
~symmetry class & $\beta \rightarrow 2$ \\
\hline
~asymptotic limit & $t \rightarrow \infty$ \\
\hline
~(asymptotic) density & ${\bar \rho} \rightarrow \log(t/2\pi)/(2\pi)$ \\
\hline
~Heisenberg time & $T_H = h {\bar \rho} \rightarrow \log(t/2 \pi)$ \\
\hline
~action & $S_p \rightarrow t \log p$ \\
\hline
~period & $T_p \rightarrow \log p$ \\
\hline
~rescaled period & $\tau_p \rightarrow \log p/\log(t/2 \pi)$ \\
\hline
~Lyapounov exponent & $\lambda_p \rightarrow 1$ \\
\hline
~stability factor & $|\det(M_{p}^r - 1)| \rightarrow p^r$ \\
\hline
~dynamical zeta & $Z(s) \rightarrow \zeta^{-1} (1 - s)$ \\
~~~~functions & $F(s) \rightarrow \prod_{r=2}^\infty 
[\zeta (r - r s)]^{(r-1)/r^2}$~ \\
\hline
~``ergodic'' zero of $Z(s)$ & $\gamma_0 \rightarrow$ pole of $\zeta(s)$ \\
\hline
~other zeros and poles & ~~~ $\rightarrow$ zeros of $\zeta (s)$ \\
~of $Z(s)$ and $F(s)$ & \\
\end{tabular}
\end{minipage}
\vspace{0.5cm}
\caption{Correspondences for the ``Riemann dynamics''. The last three entries
are discussed in the text.}
\end{table}
\end{center}

\newpage

\begin{figure}
\begin{center}
\vspace{2.5cm}
\leavevmode
\epsfysize=3.6in
\epsfbox{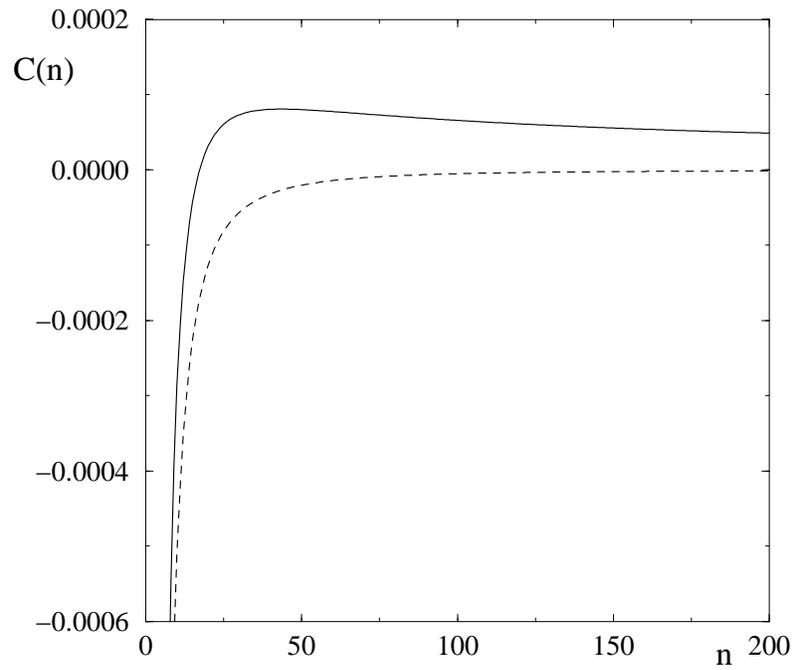}
\end{center}
\caption{Spacing autocovariances for a three-dimensional diffusive system,
with $\beta = 2$. The dashed curve is the random matrix term $C_{\rm rm}$; the
continuous curve includes the contribution $C_{\rm dif}$, Eq.(\ref{id}).}
\label{fig1}
\end{figure}

\pagebreak

\begin{figure}
\begin{center}
\leavevmode
\epsfysize=3.6in
\epsfbox{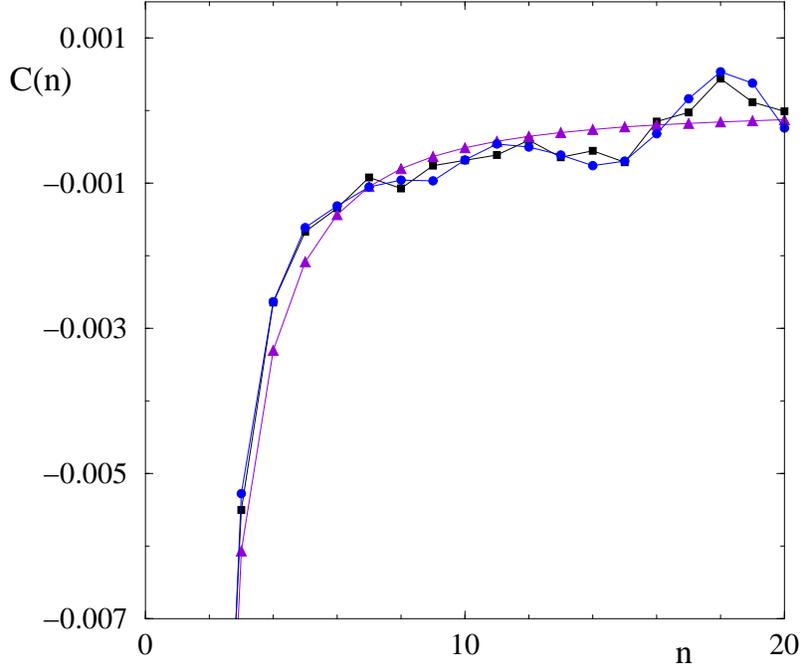}
\end{center}
\caption{ Spacing autocovariances $C(n)$ in the range $n\leq 20$. Triangles:
random matrix result, from Eq.(\ref{irmt}). Squares: data from Odlyzko computed
from Riemann zeros in an interval starting around the $10^{12}$-th zero at
$t=267653395648.8475$ and containing $50 000$ zeros. Circles: theoretical
curve, from Eq.(\ref{i3r}). Points are joined by solid lines to guide the
eye.}
\label{fig2}
\end{figure}

\pagebreak

\begin{figure}
\begin{center}
\leavevmode
\epsfysize=3.6in
\epsfbox{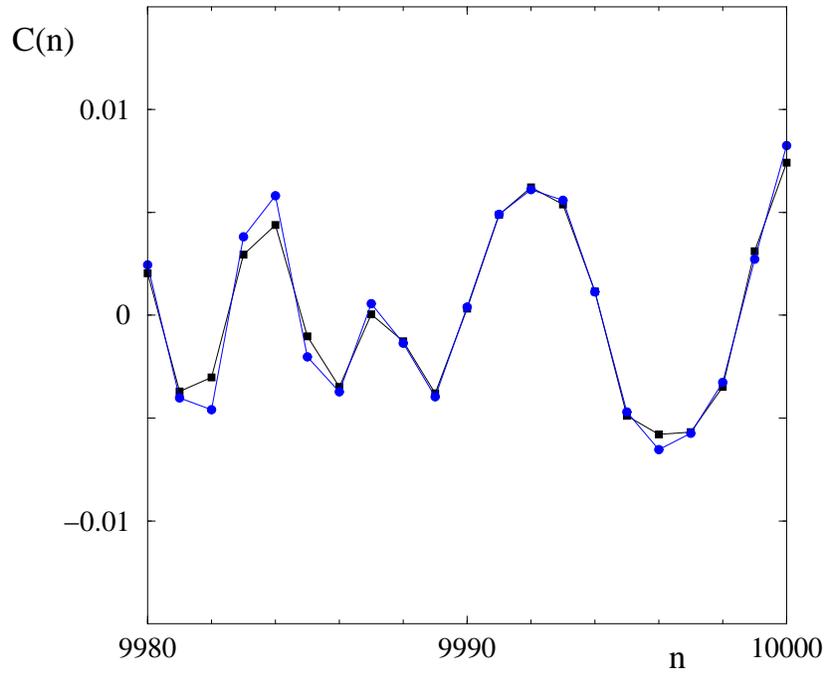}
\end{center}
\caption{Spacing autocovariances $C(n)$ for Riemann zeros in the range $9980
\leq n \leq 10000$. Squares, from data; circles, from theory, as explained in
the caption of Fig.~\ref{fig2}.}
\label{fig3}
\end{figure}

\pagebreak

\begin{figure}
\begin{center}
\leavevmode
\epsfysize=6.0in
\epsfbox{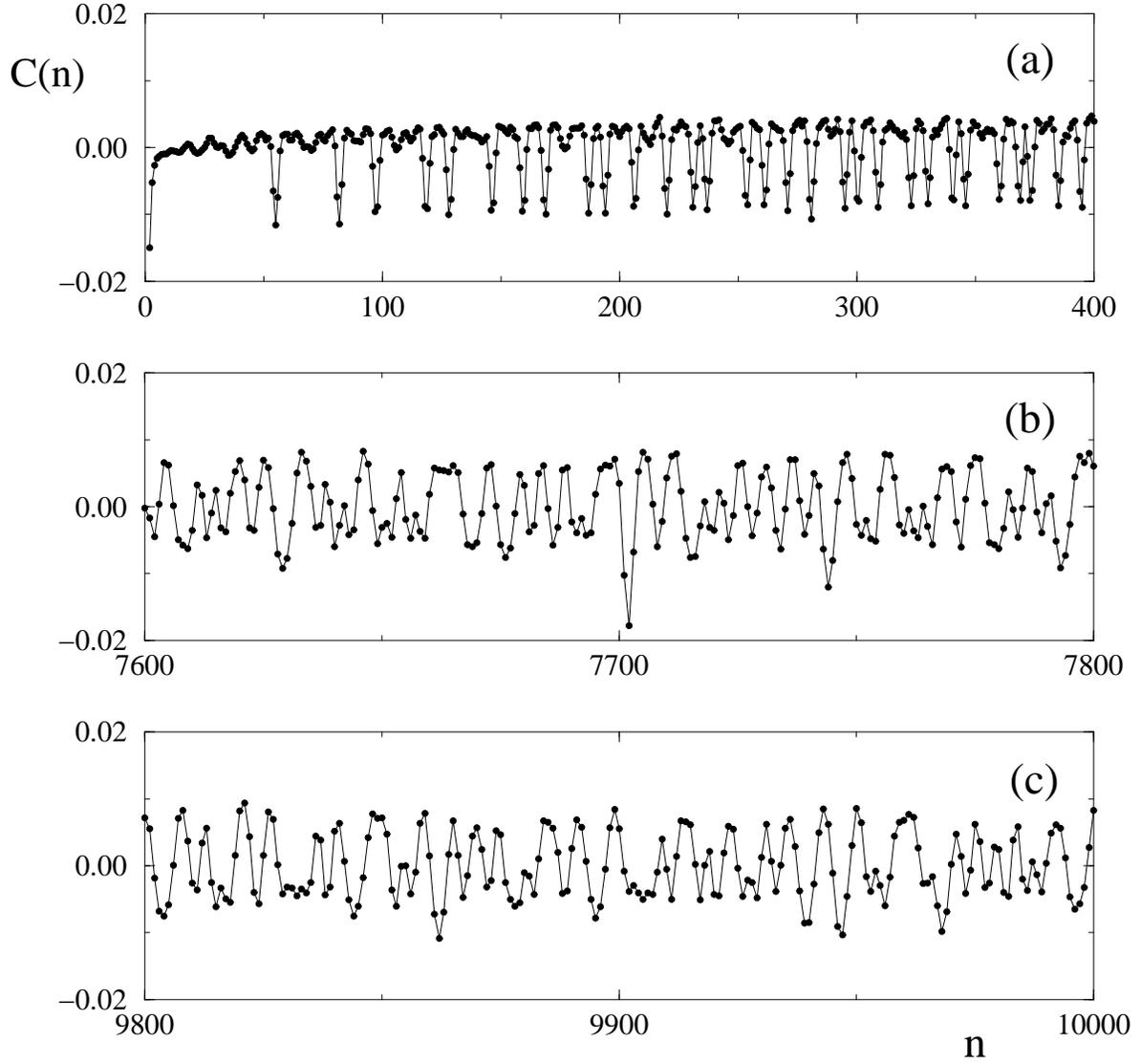}
\end{center}
\caption{Spacing autocovariances $C(n)$ for Riemann zeros for three different
ranges of $n$ to illustrate several regimes and phenomena. Circles, from
theory (Eq.(\ref{i3r}), see caption of Fig.~\ref{fig2}). The last part of
Fig.~4(c) corresponds to Fig.~\ref{fig3}.}
\label{fig4}
\end{figure}

\pagebreak

\begin{figure}
\begin{center}
\leavevmode
\epsfysize=5.8in
\epsfbox{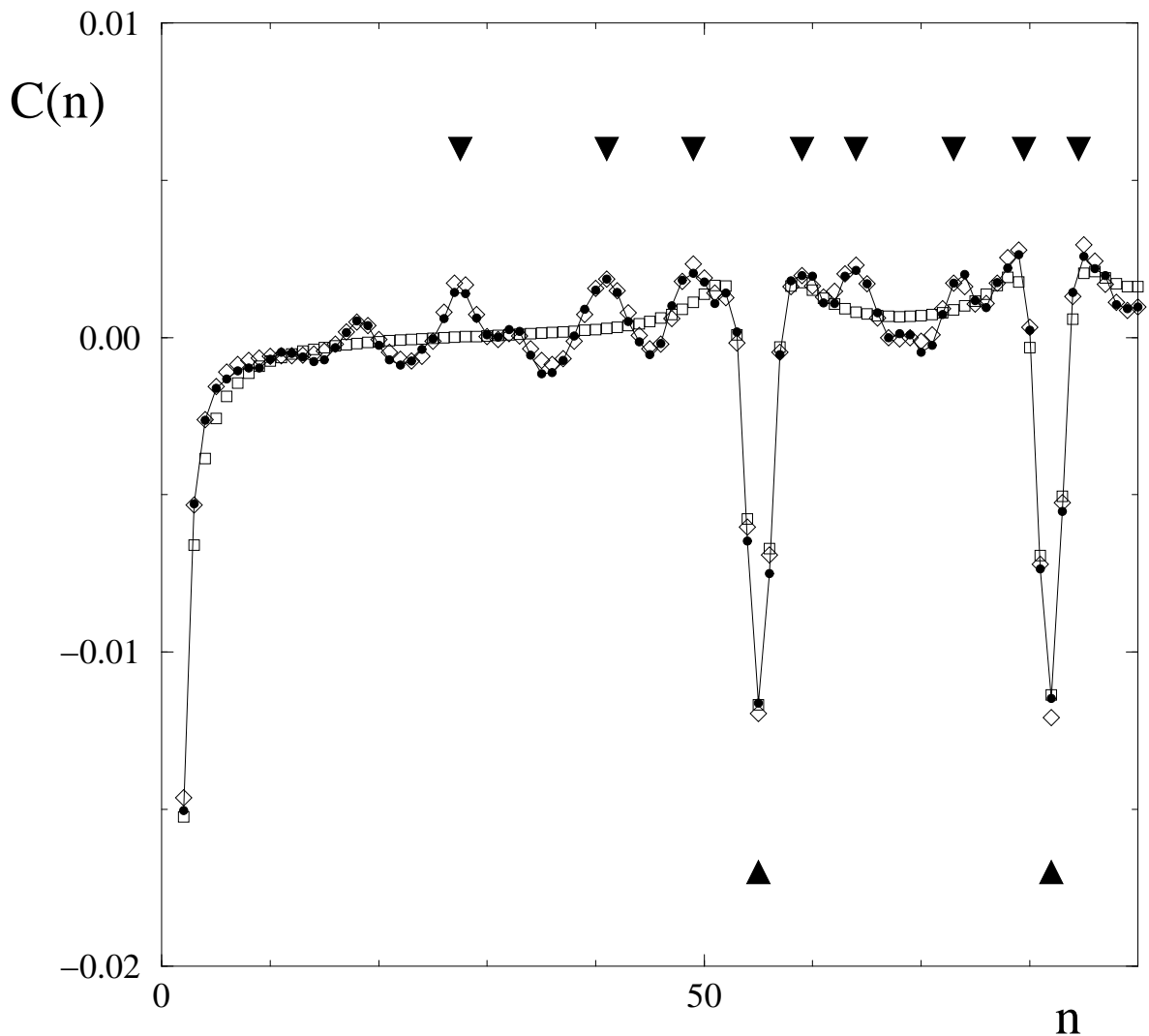}
\end{center}
\caption{Spacing autocovariances $C(n)$ for the Riemann zeros. Enlarged view
of the left part of Fig.~\ref{fig4}(a). Circles: from periodic orbit theory,
as in Figs~\ref{fig2}, \ref{fig3} and \ref{fig4}. Squares: from the resonance
formula Eq.(\ref{ipogr}) where only the $r=1$ term has been kept. Diamonds:
from Eq.(\ref{ipogr}), including up to the $r=3$ term. The two arrows in the
lower part of the figure indicate the rescaled position ${\bar \rho}\, t_\mu$
of the first two complex zeros $1/2 + i \, t_\mu$ of $\zeta (s)$. The eight in
the upper part correspond to the first sub-resonances ${\bar \rho} \, t_\mu /2$
of the first eight zeros of $\zeta (s)$. Some other (positive) peaks visible
on the figure correspond to higher order sub-resonances.}
\label{fig5}
\end{figure}

\pagebreak

\begin{figure}
\begin{center}
\leavevmode
\epsfysize=5.0in
\epsfbox{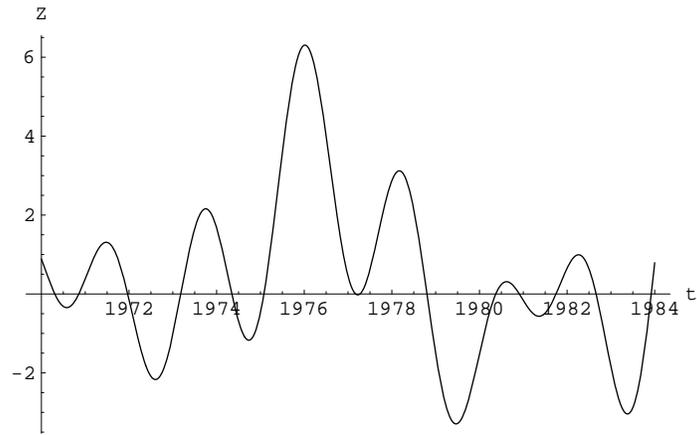}
\end{center}
\caption{The function ${\cal Z}(t)$, Eq.(\ref{hardy}), showing two
close-lying zeros of $\zeta (1/2 + i t)$ around $t=1977.2$. This almost
degeneracy is at the origin of the large peak observed in $C(n)$ at $n= {\bar
\rho} \, t = 7702$ in Fig.4(b).}
\label{fig6}
\end{figure}

\end{document}